\documentclass[twocolumn,aps,nofootinbib]{revtex4-1}

\usepackage{graphicx}
\usepackage{epstopdf}
\usepackage{latexsym}
\usepackage{amsmath}
\usepackage{amssymb}
\usepackage{color}
\usepackage{mathrsfs}

\usepackage[center]{subfigure}

\begin{document}

 \newcommand{\bq}{\begin{equation}}
 \newcommand{\eq}{\end{equation}}
 \newcommand{\bqn}{\begin{eqnarray}}
 \newcommand{\eqn}{\end{eqnarray}}
 \newcommand{\nb}{\nonumber}
 \newcommand{\lb}{\label}
 \newcommand{\be}{\begin{equation}}
\newcommand{\en}{\end{equation}}
\newcommand{\PRL}{Phys. Rev. Lett.}
\newcommand{\PL}{Phys. Lett.}
\newcommand{\PR}{Phys. Rev.}
\newcommand{\CQG}{Class. Quantum Grav.}

\title{Quasinormal Modes for Dynamical Black Holes}

\author{Kai Lin $^{a}$}\email{lk314159@hotmail.com}
\author{Yang-Yi Sun $^{a}$}\email{sunyy@cug.edu.cn}
\author{Hongsheng Zhang $^{b}$} \email{sps\_zhanghs@ujn.edu.cn}

\affiliation {a) Hubei Subsurface Multi-scale Imaging Key Laboratory, Institute of Geophysics and Geomatics, China University of Geosciences, Wuhan 430074, Hubei, China}
\affiliation {b) School of Physics and Technology, University of Jinan, 336 West Road of Nan Xinzhuang, Jinan 250022, China}

\date{\today}

\begin{abstract}
    Realistic black holes are usually dynamical, noticeable or sluggish. The Vaidya metric is a significant and tractable model for simulating a dynamical black hole. In this study, we consider scalar perturbations in a
    dynamical Vaidya black hole, and explore the quasinormal modes by employing the matrix method. We find the proper boundary conditions of the quasinormal modes from physical analysis in the background of a dynamical
    black hole for the first time. The results show that the eigenfrequencies become different at the apparent horizon and null infinity, because the physical interactions propagate with finite velocity in nature. Any
    variation of the hole does not affect the boundary condition at null infinity in a finite time. The quasinormal modes originated around the horizon would not immediately come down to, but slowly goes to the final
    state following the mass accretion process of the hole. The precision of the matrix method is quite compelling, which reveals more details of the eigenfrequencies of the quasinormal mode of perturbations in the Vaidya
    spacetime.
\end{abstract}


\maketitle

\section{Introduction}
\renewcommand{\theequation}{1.\arabic{equation}} \setcounter{equation}{0}

Observations from LIGO and Virgo\cite{LIGO1,LIGO2,LIGO3,LIGO4,LIGO5,LIGO6,LIGO7,LIGO8,LIGO9}, and Event Horizon Telescope\cite{EHT1,EHT2,EHT3,EHT4,EHT5} verify the existence of the black holes in our universe. Black holes
have different origins, such as the gravitational collapse of massive stars, the accretion or merger of compact stars, and the collapse of over-density region in the early universe. Classically, any particle or radiation inside the horizon can never escape beyond it. Therefore an observer outside the horizon is ignorant about the interior of a black hole. Fortunately, we can study the properties of black holes through several circumstantial
phenomena, such as the surrounding accretion processes, gravitational wave radiation, and gravitational lens. Due to the complexity of the astronomical environment, people hope to receive a characteristic signal of a black
hole to facilitate its location and to have an insight into its properties. For this purpose, it is intriguing for us to explore the Quasinormal modes (QNMs) process associated with a black hole.

A linear perturbation of a black hole is a critical tool to reveal the stability of the black hole. An unstable perturbation is a self-excitation mode, that leads the hole to explode eventually. In contrast the black hole would
remain if the perturbation modes are stable. The vibrations of the black hole decays gradually due to the loss of energy from gravitational radiation. The process includes three stages: First of all, in the initial
perturbation stage, the initial perturbation mainly determines the vibration behavior that hardly depends on the nature of black hole itself; The second stage is the QNMs vibration stage: The waveform can be described by a
complex frequency $\omega$, which includes a positive real part (indicating the frequency of vibration) and a negative imaginary part (indicating the rate of decrease in vibration amplitude). This complex frequency is
determined by the nature of the black hole itself. Therefore, it is just the characteristic frequency of the black hole. We can investigate the black hole according to the QNMs frequency, and even to probe the relevant
parameters in  quantum gravity through black hole physics; In the final stage, the amplitude of the vibration decreases rapidly after the QNMs perturbation phase, which is named as the late time tails. The nature of this
phase is also determined by the black hole itself.

On the other hand, the gravitational wave from the binary compact objects also shows that the gravitational wave radiation in this process includes three phases: Inspiral, Merger, and Ringdown. Linear perturbation of a black
hole can describe the waveform of the ringdown phase. The ability of the current gravitational wave detectors is insufficient to preciously show the waveform of the ringdown phase. Fortunately, the next generation of the ground
gravitational wave detectors could allow us to survey the details of this phase \cite{Ringdown1,Ringdown2,Ringdown3,Ringdown4,Ringdown5,Ringdown6,Ringdown7,Ringdown8,Ringdown9,Ringdown10}. Therefore, investigating QNMs will
shed light on the black hole physics and gravitational wave astronomy in the future.

So far, people proposed various methods to study the QNMs of black holes, such as the WKB method\cite{WKB1,WKB2,WKB3,WKB4,WKB5,WKB6,Konoplya1}, the asymptotic iteration method (AIM)\cite{AIM1,AIM2,AIM3,AIM4}, and other
methods \cite{Expand1,NewM1,NewM2,NewM3,NewM4}. However, almost all of them are designed for studying the case of static or stationary black holes. Regarding to the realistic black holes in astrophysics, their mass may either
decrease in evaporation due to the quantum Hawking radiation effect, or increase due to the accretion process. There is no doubt that the dynamical evolution plays an essential role in astronomical observation and
theoretical research of black holes. However, few studies pay attention to the QNMs of the dynamical black hole. In general, obtaining an analytical solution from the black hole perturbation equation is quite challenging.
The situation becomes more complicated for a dynamical black hole. A few numerical methods are recommended to resolve the time-dependent partial differential equations\cite{Hod1,Wang1,Wang2,Elcio1,Cecilia1}. An available
method to calculate the QNMs in  dynamical spacetimes is the finite difference method invoked by Wang and Abdalla et al. in \cite{Wang1,Wang2,Elcio1,Cecilia1}. The method fits the frequency from the curve of the black hole
perturbation waveform. It is a straightforward way to calculate the QNMs frequency. The  method precision in certain situations leaves much to be desired. Especially for a small angular momentum $L$, the QNMs phase is
almost unobservable in the curve of the perturbation waveform.

In the previous investigations, we propose the matrix method to calculate the QNMs of black holes\cite{Lin1,Lin2,Lin3}. The results illustrate that the matrix method can be widely utilized to investigate the asymptotical
flat and (Anti-)de Sitter black hole space-times. This approach   gives us an insight into the QNMs frequency and eigenfunction of the accreting Vaidya-AdS black hole \cite{Lin4}, in which we considered the case with mass
increasing from an initial mass $M_i$ to another final mass $M_f$ in finite time. Therefore, the QNMs frequency $\omega_i$ and its eigenfunction in the Schwarzschild black hole spacetime with mass $M_i$ is naturally assumed
as the initial condition. By applying the finite difference method along the time, we solved the QNMs equation in the dynamical black hole spacetime. The results showed that the QNMs frequency fails to reach the final state
(QNMs frequency of Schwarzschild with mass $M_f$) immediately, but approaching this value after a long time.

In this study, we examine the QNMs of Vaidya black hole by utilizing the matrix method. In such an asymptotically flat dynamical spacetime, it is essential to consider the boundary condition being more complicated than that
of the AdS spacetime. The reason is that the wave function of the AdS spacetime vanishes always at infinity due to the divergence of the potential function. However, the wave function remains finite in an asymptotically
flat spacetime. Furthermore, the boundary conditions at infinity cannot be changed in a finite time due to the limited velocity of physical interactions. Therefore, the QNMs frequency of the Vaidya black hole should
depend on both the radial spatial and the temporal coordinates. For the reason mentioned above, a suitable boundary condition is required. In the next section, we illustrate the numerical calculation process of the
dynamical Vaidya black hole using the matrix method. Section III shows the numerical results. We conclude the investigation in section IV.

\section{Scalar Perturbation and Quasinormal Modes of Vaidya Black Hole}
\renewcommand{\theequation}{2.\arabic{equation}} \setcounter{equation}{0}

Typically, the astronomical environment for a black hole is rather complicated. Usually, a black hole lives in an accretion disk, surrounded by gases and dark matters. Actually, in the history of the universe, almost all the
realistic black holes are in accretion, getting larger and larger. Some particular black holes may be undergoing the Penrose process to release energy. Even for a hole in the clean astronomical environment, it is not
completely static due to the quantum evaporation, i.e., Hawking radiation. It is a formidable task to find an analytical solution for the realistic accreting black hole. Fortunately, Vaidya suggested a dynamical metric
that is surrounded by null dust. Vaidya called it a shining star, as it radiates null matters described by retarded Eddington coordinates. It also describes absorbing stars by the advanced Eddington coordinates. The Vaidya
metric is an appropriate theoretical model for dynamical spherical black holes. It is extended to various modified gravities, for example, the massive gravity \cite{self1}.
For an accretion black hole, we write the Vaidya metric in the advanced Eddington coordinates as
\bqn
\lb{metric1}
ds^2=-f(v,r)dv^2+2drdv+r^2\left(d\theta^2+\sin^2\theta d\varphi^2\right),
\eqn
with
\bqn
\lb{metric2}
f(v,r)=1-\frac{2M(v)}{r}\equiv1-\frac{r_0(v)}{r},
\eqn
where $r_0=r_0(v)$ is the apparent horizon, which satisfies $f(v,r_0(v))=0$.

Substituting above metric into the massless scalar field equation $\partial_\mu\left(\sqrt{-g}g^{\mu\nu}\partial_\nu\Phi\right)=0$, we get the radial scalar perturbation equation as follows
\bqn
\lb{scalar1}
f\partial_r(f\partial_r\phi)+2f\partial_v\partial_r\phi-V\phi=0,
\eqn
where the potential term $V=\frac{f}{r^2}\left[r\partial_rf+L(L+1)\right]$ and the field function comes from $\Phi=\sum \frac{\phi(v,r)}{r}Y\left(\theta,\varphi\right)$.

In order to simplify the calculation, we do the transformation $r\rightarrow \frac{r}{r_i}$ and $v\rightarrow \frac{v}{r_i}$, so that the initial value of apparent horizon $r_i$ becomes 1.

 Before making further separation of variables, we explore the physical significance of the coordinate $v$. For static spacetime, there is a natural 3+1 decomposition since $\frac{\partial}{\partial t}$ is Killing vector. And
 thus, an observer who shifts along the integral curve of this vector senses an invariant spacelike 3-surface. This invariance leads to a conserved current $J$, which is defined as,
 \be
 J_a=T_{ab}\left(\frac{\partial}{\partial t}\right)^b.
 \en
 Here $T_{ab}$ denotes the stress energy of a field, or only a conserved symmetric tensor on the manifold, which satisfies $\nabla^a T_{ab}=0$. Then it is easy to demonstrate that $J$ is a conserved current,
 \be
 \nabla^a J_a=0.
 \label{Ja}
 \en
 The conserved charge $E$ corresponding to this current reads
 \be
 E=-\int *J^{\sim},
 \label{E=}
 \en
 where $*$ marks the Hodge dual of a form, and ${}^{\sim}$ labels a restriction to the spacelike 3-space, i.e., the hypersurface orthogonal to $\frac{\partial}{\partial t}$.
 If we invoke the stress energy of the spacetime, which expands the spacetime geometry according to the Einstein equation, we directly arrive at the Misner-Sharp energy as a conserved charge.
 This argument can be extended to higher dimensions and modified gravities \cite{self1, self2, self3, self4,self5}. For a probe field (no back reaction to the metric), one also directly defines
 the conserved current and conserved charge corresponding to its stress energy by using (\ref{Ja}) and ( \ref{E=}).

 Generally, for a dynamical spacetime, no time-like field has priority to decompose the spacetime. Fortunately, for a dynamical spacetime in spherical symmetry we have a natural extension of the Killing vector, named
  after Kodama \cite{Kodama}. Careful investigations imply that the Kodama vector is the one with priority to decompose dynamical spacetimes in spherical symmetry \cite{KodamaT,Kodama1,Kodama2}. The resulted time coordinate is also called
  Kodama time. The key property of Kodama vector in spherical symmetry is,
 \be
 \nabla_{(a} K_{b)}=0,
 \en
 which exactly follows the property of the Killing vector. Thus, naturally, we construct a conserved current $J$ from a conserved stress energy $T_{ab}$,
 \be
 J_a=T_{ab}K^b.
 \en
 Similar to the static case, the conserved charge corresponding to this current reads,
 \be
 E=-\int *J^{\sim}.
 \label{1E=}
 \en

 For the dynamical spherical metric (\ref{metric1}) in areal coordinate,  the Kodama vector reads,
  \be
  K^b=\left(\frac{\partial}{\partial v}\right)^b.
  \en
  For the stress energy of the spacetime metric, the conserved current becomes,
  \be
  J=\frac{1}{8\pi}\left((-1+f+rf')\frac{dr}{r^2}+(f-f^2-rff'-r\dot{f})\frac{dv}{r^2}\right),
  \en
 where a prime denotes derivative with respect to $r$, and a dot for $v$. The Hodge dual of $J$ reads,
 \be
 *J=\frac{\sin\theta}{8\pi} \left(  (-1+f+rf')dr\wedge d\theta \wedge d\phi+f\dot{f} dv\wedge d\theta \wedge d\phi\right) .
 \en
 Thus a restriction of $*J$ to the spacelike hypersurface orthogonal to $K$ is,
 \be
 *J^{\sim}=\frac{1}{8\pi}  (1-f-rf')dr\wedge d\cos\theta \wedge d\phi.
 \en
 The resulted conserved charge is,
 \be
 E=-\int *J^{\sim}=\frac{r}{2}(1-f).
 \en
 This is a new definition of Misner-Sharp energy through conserved charge method \cite{self1}. One easily confirms that the conserved charge is exactly the mass parameter $M(v)$ for Vaidya metric. In a word, $v$ is proper
 time coordinate, and corresponds to a conserved charge (energy) in dynamical spherical symmetry.

   Thus, it is reasonable to write the scalar field as $\phi\sim e^{-i\omega(v) v}$. At the apparent horizon, the eigen Kodama energy is directly proportional to $\omega+v\partial_v{\omega}$. Comparing to the case of static
   metric, a mode $e^{-i\omega(v) v}$ does not imply a constant eigenenergy. This is not surprised since the background spacetime is dynamical. The energy of a particle is inherently variable when shifting along
   the vector field which   carries the intrinsic symmetry, i.e., the Kodama field. The point is that the stress energy of the scalar field yields a conserved current and a conserved charge (energy) with the aids of the Kodama
   vector.

  Mass accretion shifts the position of apparent horizon $r_0=r_0(v)$ immediately. In detail, at the apparent horizon, the potential term vanishes, so that perturbation equation reduces as
\bqn
\lb{scalar2}
f\partial_r(f\partial_r\phi)+2f\partial_v\partial_r\phi=0,
\eqn
and the above equation permits two solutions: the ingoing mode $\phi_1=\tilde{C}_1e^{-i\omega(v)v}$ ($\tilde{C}_1$ is a constant and $\omega(v)$ is the temporal dependent frequency), and the outgoing mode $\phi_2$ that
satisfies $f\partial_r\phi_2+2\partial_v{\phi_2}=0$. The boundary condition requires the outgoing mode being vanished, because nothing can escape from the classical black hole.

However, the change of the apparent horizon would not affect the boundary condition at infinity in a finite time, because the speed of light is  finite. In this paper, we consider that the initial status of mass is a
constant, so the metric becomes static Schwarzschild solution, and boundary solution at infinity is the outgoing mode $\phi=\tilde{C}_0e^{-i\omega_i(v-2r)}r^{2i\omega_ir_i}$, where $\tilde{C}_0$ is a constant and $\omega_i$
is the QNMs frequency of the initial Schwarzschild status as $v\le v_a$ in (\ref{mass1}).

Therefore, the field function in the master equation can be recomposed as
\bqn
\lb{phi1}
\phi=e^{-i\left[\omega_i+\frac{r_0(v)}{r}\left(\omega(v)-\omega_i\right)\right]v}e^{2i\omega_ir}r^{2i\omega_ir_i}\Psi(v,r),
\eqn
so that the function $\Psi$ satisfies the boundary condition
\bqn
\lb{phi2}
\Psi=\left\{
  \begin{array}{ll}
    C_1 & r\rightarrow r_0(v)\\
    C_0 & r\rightarrow \infty\\
  \end{array}
\right.
\eqn
where we can find that the constants $C_1$ and $C_0$ are not independent, and $C_0$ can determine $C_1$ by the calculating of the eigenvalues and eigenvectors.
Above discussion and Eq.(\ref{phi1}) imply that the QNMs frequencies become different at different radial positions $r$. The effect also comes from the fact that the speed of physical interactions propagate is finite. Thus any
variance around the horizon does not affect the physical environment at finite distance. After the change of horizon, the QNMs frequency $\omega(v)$ at horizon in Eq.(\ref{phi1}) will change immediately, but $\omega_i$ at infinity won't change in a finite time.

We also introduce a new variable $x\equiv1-\frac{r_0(v)}{r}$, so that the range $0\le x\le 1$ ($x=0$ at horizon and $x=1$ at infinity). Under the above transformations, the master equation becomes
\bqn
\lb{scalar3}
&&H_{xx}\Psi''+H_{xv}\dot{\Psi}'+H_x\Psi'+H_v\dot{\Psi}+H_0\Psi=0,\nb\\
&&H_{xx}=(x-1)^3[2\dot{r}_0-x(1-x)],\nb\\
&&H_{xv}=2(1-x)^2r_0,\nb\\
&&H_x=i(1-x)\left\{-2 r_0(v) \left[2 \omega _i \dot{r}_0+(x-1) \left(x \omega _i\right.\right.\right.\nb\\
&&~~~~~\left.\left.+v (x-1) \dot{\omega}\right)+(x-1)^2 \omega\right]\nb\\
&&~~~~~+(x-1) \left[\dot{r}_0 \left(4 \omega _i \left(r_i+v
   (x-1)\right)+2 i\right)\right.\nb\\
&&~~~~~+(x-1) \left(x \left(4 r_i \omega _i+2 v (x-1) \omega _i+3 i\right)-i\right)\nb\\
&&~~~~~\left.\left.-2 v (x-1) \omega \left(2 \dot{r}_0+(x-1) x\right)\right]\right\},\nb\\
&&H_v=2ir_0\left[2\omega_ir_0\right.\nb\\
&&~~~~~\left.+(1-x)(2\omega_ir_i-(1-x)v(\omega_i-\omega))\right],\nb\\
&&H_0=(1-x) \left\{(x-1) \left[2 v \left(\omega_i-\omega\right) \dot{r}_0 \left(2 r_i \omega_i\right.\right.\right.\nb\\
&&~~~~~\left.+v (x-1) \omega_i+(v-v x) \omega+i\right)\nb\\
&&~~~~~+x \omega_i^2 \left(2 r_i+v(x-1)\right)^2+i \omega_i \left((4 x-2) r_i\right.\nb\\
&&~~~~~\left.+v x (3 x-4)+v\right)+v (x-1) \omega \left(-x \left(4 r_i \omega_i\right.\right.\nb\\
&&~~~~~\left.\left.+2 v (x-1) \omega _i+3 i\right)+v (x-1) x \omega+i\right)\nb\\
&&~~~~~\left.+L^2+L-x+1\right]+2 r_0 \left[\omega \left(2 v \omega _i \dot{r}_0\right.\right.\nb\\
&&~~~~~+(x-1) \left(-2 r_i \omega_i+v \omega_i+v^2 (x-1) \dot{\omega}\right.\nb\\
&&~~~~~\left.\left.-i\right)\right)-v (x-1) \dot{\omega} \left(2 r_i \omega_i+v (x-1) \omega_i\right.\nb\\
&&~~~~~\left.+i\right)+\omega_i^2 \left(-\left(2 x r_i+2 v \dot{r}_0+v (x-1) x\right)\right)\nb\\
&&~~~~~\left.\left.+v(x-1)^2 \omega^2\right]+4 \omega _i r_0^2 \left(v \dot{\omega}+\omega\right)\right\}
\eqn
where  ${\cal F}'\equiv\partial_x{\cal F}$ and $\dot{\cal F}\equiv\partial_v{\cal F}$. When $r_0\rightarrow r_i$, the spacetime reduces to Schwarzschild case with $\omega\rightarrow\omega_i$ and $\Psi\rightarrow\Psi_i$, so
the master equation becomes
\bqn
\lb{scalar4}
&&x(1-x)^2\Psi_i''+\left[(4x-1)(2i\omega_i r_i-1)\right.\nb\\
&&\left.+(3-4i\omega_i r_i)x^2\right]\Psi_i'+\left[8\omega_i^2r_i^2+4i\omega_i r_i-1\right.\nb\\
&&\left.-L-L^2-x(i+2\omega_i r_i)^2\right]\Psi_i=0,
\eqn
and we can get the initial conditions from the above equations.

To make the study concise and definite, we assume the mass accretion as follows
\bqn
\lb{mass1}
&&r_0(v)\equiv 2M(v)=\nb\\
&&\left\{
  \begin{array}{cc}
    r_i & 0\le v<v_a\\
    r_i+A+A\sin\left(\alpha(v-v_a)-\frac{\pi}{2}\right) & v_a\le v<v_a+\frac{\pi}{\alpha}\\
    r_i+2A & v_a+\frac{\pi}{\alpha}\le v\\
  \end{array}
\right.\nb\\
\eqn
In principle, $M(v)$ is a $C^2$ function. Here $M(v)$ denotes a solution which describes a spacetime initiates from a Schwarzschild one, and then accretes to some degree, and returns to a static one at last. The
 evolution of $M(v)$ is shown in fig. 1.
 \begin{figure}[tbp]
\centering
\includegraphics[width=1\columnwidth]{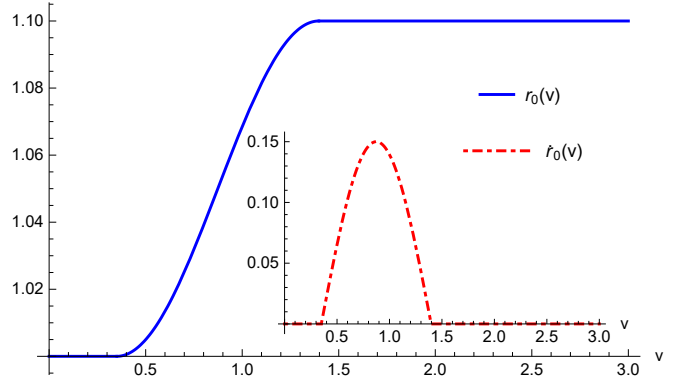}
\caption{The time dependent function for $r_0=r_0(v)$ and $\dot{r}_0$ with $A=1/20$, $\alpha=3$ and $v_a=7/20$. The initial value of $r_0$ is 1 as $v<v_a$, and it becomes 1.1 as $v>v_b$.}
\lb{Fig1}
\end{figure}
 To explicitly display the position and propagation of the QNMs, fig. 2. displays a conceptual Penrose diagram for an accretion Vaidya black hole.

 \begin{figure}[tbp]
\centering
\includegraphics[width=0.6\columnwidth]{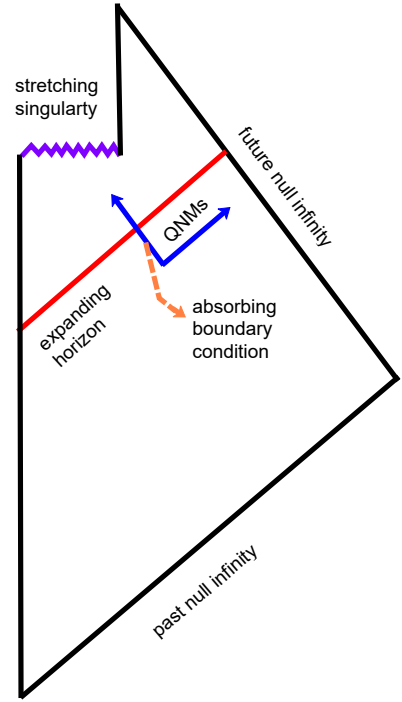}
\caption{Penrose diagram for an accreting Vaidya black hole. The horizon is expanding, while the boundary condition at future null infinity is not affected.}
\lb{penrose}
\end{figure}

In the next section, we will calculate the time-dependent QNMs frequency $\omega(v)$ by using the matrix method.

\section{Numerical Results}
\renewcommand{\theequation}{2.\arabic{equation}} \setcounter{equation}{0}

The situation we faced in the current stage is the master equation (\ref{scalar3}) with the boundary (\ref{phi2}) and initial conditions $\omega_i$ and $\Psi_i$, which satisfy the Schwarzschild black hole's QNMs
equation (\ref{scalar4}). To resolve the problem, we need discretize the partial differential equation (\ref{scalar3}) and initial equation (\ref{scalar4}). By using the matrix method, the eigenvalue
$\omega_i$ is obtained efficiently. Substituting $\omega_i$ into the matrix equation from (\ref{scalar4}) can calculate the eigenvector $\Psi_i(x_i)$ on the grid. Next, from the discrete master equations, the eigenvalue
$\omega(v_{k+1})$ and the eigenvector $\Psi(v_{k+1},x_i)$ at time $v=v_i+(k+1)\Delta v$ can be derived by $\omega(v_k)$ and $\Psi(v_k,x_i)$ at time $v=v_i+k\Delta v$, where $\Delta v$ is the step of time coordinate. We show
the calculation results in figs.3 and 4.

\begin{figure*}[tbp]
\centering
\includegraphics[width=0.9\columnwidth]{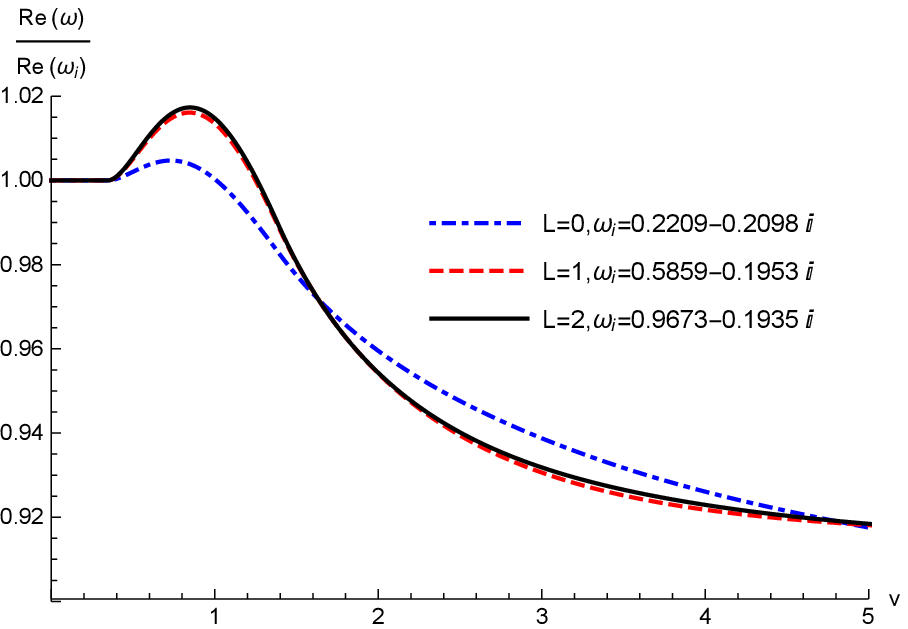}\includegraphics[width=0.9\columnwidth]{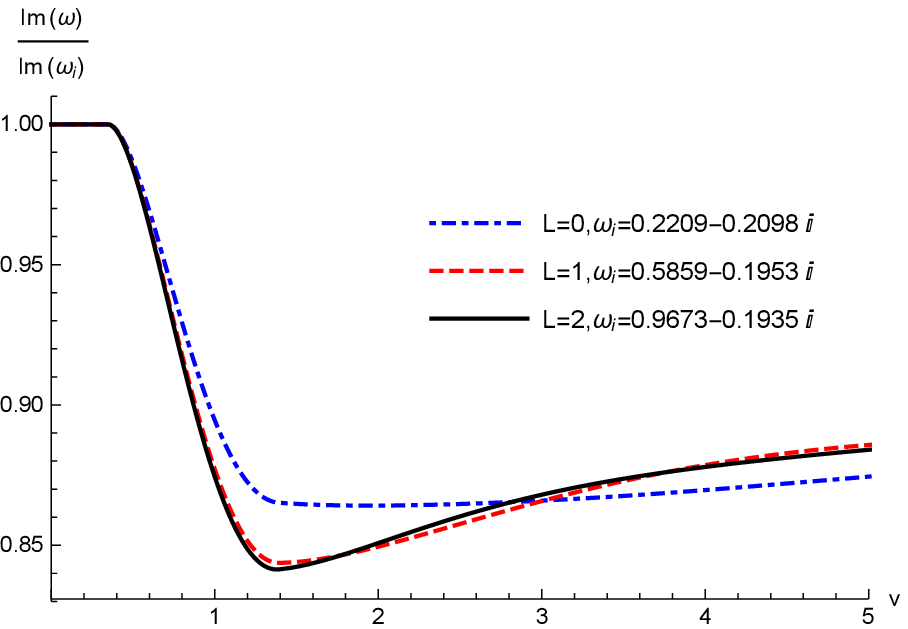}
\caption{The time dependent function for QNMs frequency $\omega$ with $L=0,1,2$ and the step $\Delta x=1/21$.}
\lb{Fig2}
\end{figure*}

\begin{figure*}[tbp]
\centering
\includegraphics[width=0.9\columnwidth]{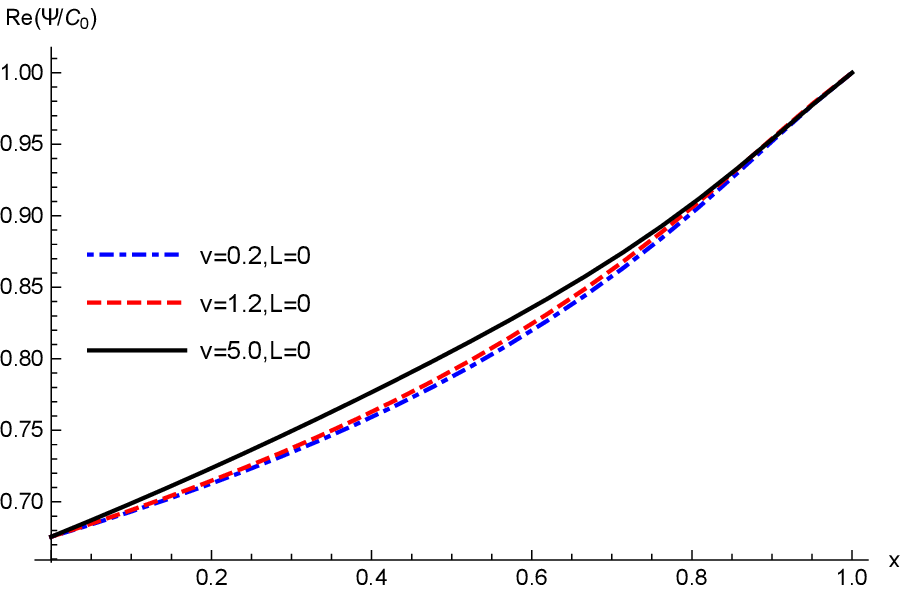}\includegraphics[width=0.9\columnwidth]{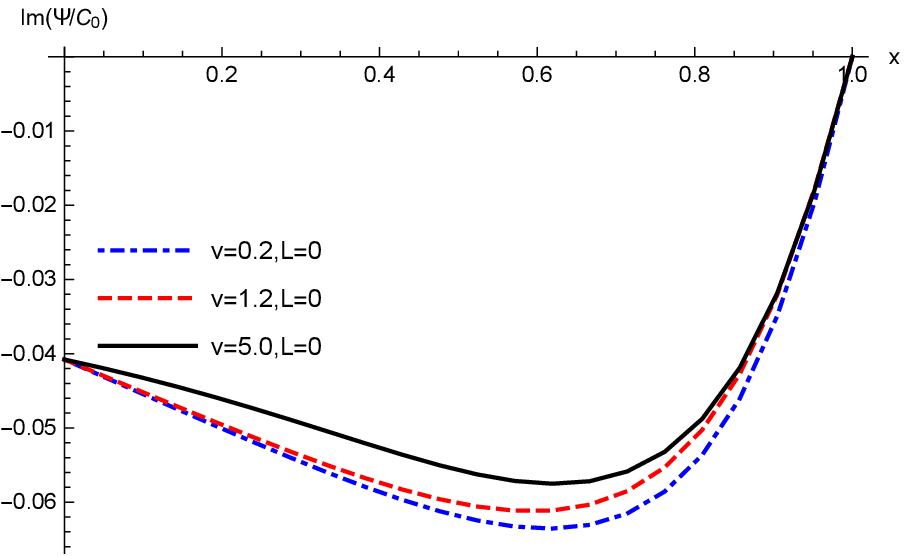}
\includegraphics[width=0.9\columnwidth]{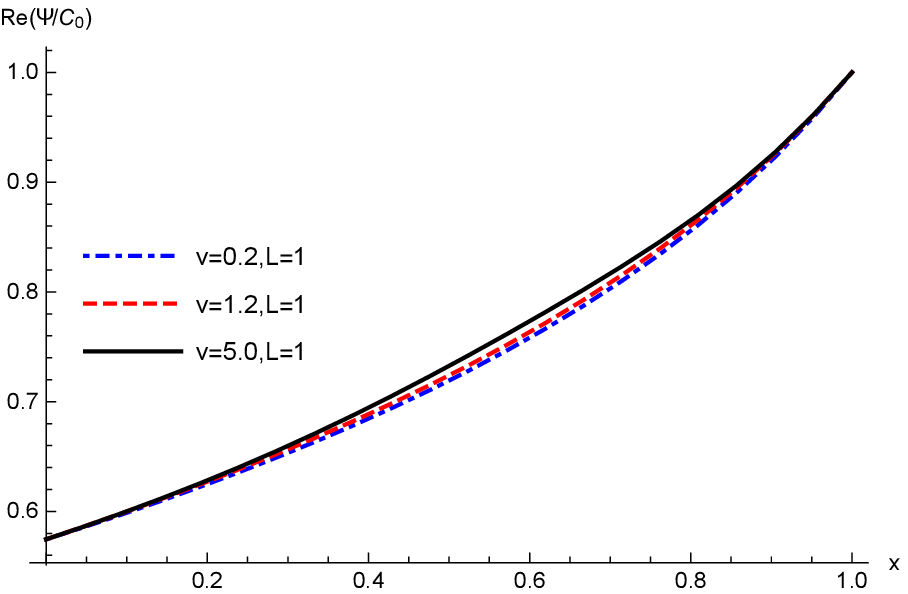}\includegraphics[width=0.9\columnwidth]{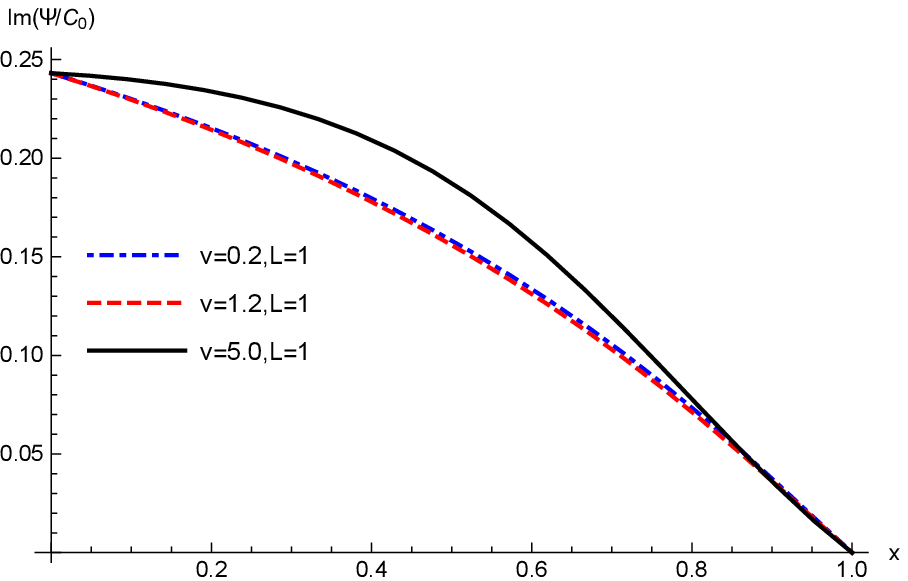}
\includegraphics[width=0.9\columnwidth]{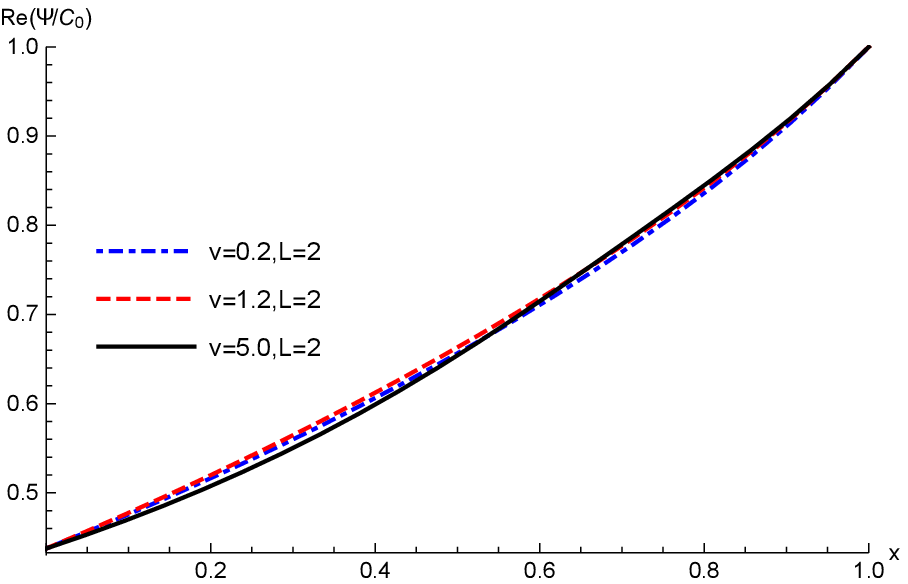}\includegraphics[width=0.9\columnwidth]{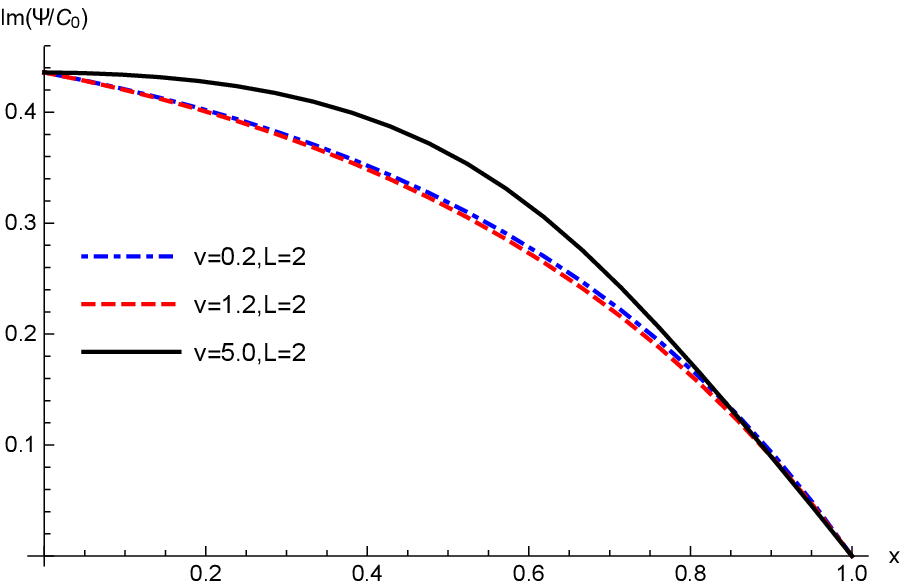}
\caption{The $x=1-\frac{r_0(v)}{r}$ dependent function for function $\Psi$ with different $v$ with $L=0,1,2$ and the step $\Delta x=1/21$.}
\lb{Fig3}
\end{figure*}

 Above figures show the properties of QNMs frequency at the apparent horizon of an accreting Vaidya black hole. Comparing fig.1 with fig.2, we find that $\omega$ changes immediately after the horizon $r_0$ varies from the initial position. Since $\omega(v)$ is the QNMs frequency at the horizon, the variance of $r_0$ can affect $\omega(v)$ soon. However, on the other hand, after the horizon $r_0$ achieves at final size (namely Vaidya spacetime terminated at a static Schwarzschild spacetime with larger horizon), we find the frequency $\omega$ can not cease variance at once, but $\omega$ slowly goes to the value of QNMs frequency of final state Schwarzschild spacetime. The
 phenomenon can be found in QNMs of dynamical black holes\cite{Wang1,Wang2,Elcio1,Cecilia1,Lin4}, and is helpful to differentiate signals from dynamical black holes and static ones.

 Fig.3 displays the trend of real and imaginary parts with respect to the variation of angular momentum $L$. A mode with larger angular momentum $L$ implies a larger real part of $\omega$, but a smaller absolute value of its imaginary part. Another property displayed in fig.3 is that the frequency $\omega(v)$ is not monotonic. Real part of $\omega(v)$ hits a maximal value and then decreases to the final state value gradually, while the imaginary part of $\omega(v)$ falls to a minimal value and then increases to the final value slowly. Interestingly, the  maximal value real part appears of $\omega(v)$ near the maximal value of $\dot{r}(v)$, as a contrast, the minimal value of imaginary part is near the position where $r_0$ reaches the final state, as shown in fig.1. What's more, for larger $L$, the above property is more evident, that is, the maximal value of real part becomes larger, and the minimal value of imaginary part becomes smaller.

Fig.4 shows the variance of $\Psi$, which is an eigenfunction of QNMs problem. We find that, at different time $v$, the imaginary part of $\Psi$  changes visibly. Simultaneously, its real part does not change drastically, and
the variance becomes milder for a larger $L$.

\section{Conclusion}
\renewcommand{\theequation}{4.\arabic{equation}} \setcounter{equation}{0}

In this study, we apply the matrix method and the time difference technology to study the Quasinormal modes (QNMs) process of the dynamic black holes. Because this method straightforwardly penetrates the eigenfrequency
of QNMs, the numerical results are more accurate than those obtained by the finite difference method. On the other hand, due to the fact that the speed of physical interaction is limited, the change in the mass of the black
hole would not affect the boundary conditions at infinity in finite time. This implies that the frequency of QNMs of dynamic black holes is a function of time and space. There are similar results in the study of
gravitational collapse: the information of gravitational collapse cannot reach infinity in a finite time \cite{Weinberg}.

Since a black hole is a simple celestial body with strong gravitation, people tend to treat a black hole as a probe to explore quantum gravity in reasoning. The status of black holes in gravitational theory may be
similar to hydrogen atoms in quantum theory. Without considering the changes in atomic energy levels, it is impossible to conduct in-depth research on quantum theory. Similarly, the study of dynamical black holes will
inevitably have a profound impact on quantum gravity theory and astronomy. We will continue to study and discuss the open problems, such as bound and scattering problems, in the future.

\begin{acknowledgments}
We gratefully acknowledge the financial support from
National Natural Science Foundation of China (NNSFC) under contract No. 11805166 and 41804154,
National Key R\&D Program of China (2020YFC2201400), and Shandong Province Natural Science Foundation (ZR201709220395).
\end{acknowledgments}

\end{document}